\documentclass[a4paper,UKenglish,cleveref, autoref, thm-restate]{lipics-v2021}
\usepackage{subcaption}
\usepackage{tcolorbox}
\title{What Makes Software Issue Resolution Tasks Difficult for Agents?}

\author{Ebtesam Al-Haque}{Department of Computer Science, George Mason University, Fairfax, VA }{ehaque4@gmu.edu}{https://orcid.org/0009-0005-1992-7193}{}

\author{Brittany Johnson}{Department of Computer Science, George Mason University, Fairfax, VA}{johnsonb@gmu.edu}{https://orcid.org/0000-0002-0271-9647}{}

\authorrunning{E Al-Haque and B Johnson}
\Copyright{Ebtesam Al Haque and Brittany Johnson}



\category{Technical Track Paper}

\relatedversion{}

\nolinenumbers

\EventEditors{Robert Feldt, Maria Paasivaara, Daniel Mendez, Stefan Wagner, and Marvin Mu\~{n}oz Bar\'{o}n}
\EventNoEds{5}
\EventLongTitle{20th International Symposium on Empirical Software Engineering and Measurement (ESEM 2026)}
\EventShortTitle{ESEM 2026}
\EventAcronym{ESEM}
\EventYear{2026}
\EventDate{October 8--9, 2026}
\EventLocation{Munich, Germany}
\EventLogo{}
\SeriesVolume{394}
\ArticleNo{3}

\begin{document}

\maketitle

\begin{abstract}
 \textbf{Background. } Advances in agentic systems are simultaneously, and rapidly, saturating benchmarks. Despite this often discussed phenomena,  benchmark scores remain difficult to interpret due to  the lack of control and characterization of task difficulty. 
 More specifically, we currently have little understanding of what makes one task harder than another, and to what extent task difficulty is predictable from static task properties. 
 \textbf{Aims.} We propose a measurement framework to investigate and systematically quantify what structural properties of software tasks correspond to agent success rates for issue resolution tasks.
 \textbf{Method.} We conducted a large scale empirical study on CoderForge-Preview, the largest open dataset of coding agent trajectories to date, by extracting features across task patch, repository and prompt. 
 We evaluated the predictive power of each feature against task outcomes using ensemble methods, SHAP attribution, and effect size analysis.
 \textbf{Results} We found that task difficulty is substantially predictable from static features ($AUC=0.863$) and is largely driven by patch fragmentation and repository scale. Prompt linguistic features become visible among top contributors for tasks in the mid-band, revealing a layered structure of difficulty.
 \textbf{Conclusion.} The difficulty of an issue resolution task is encoded in its structure. This enables static, pre-hoc difficulty estimation and lays the groundwork for difficulty-controlled benchmark construction for evaluation of agents.
\end{abstract}

\section{Introduction}
AI-based coding agents are rapidly advancing, with the best open-source systems now resolving more than half the tasks on SWE-bench Verified~\cite{openai2024verified}. 
The aggregate solve rate, however, provides no account of \emph{why} a task is hard. Understanding this in agent outcomes is critical as agentic decision-making systems become more widely deployed.
Prior work has invested heavily in building larger and more diverse task collections~\cite{jain2025r2e,yang2026swe,badertdinov2026swe} and in validating the correctness of existing ones~\cite{aleithan2024swe,openai2024verified}, but has not systematically characterized task difficulty from static properties.
Item Response Theory (IRT)~\cite{lalor2016building} offers a framework for difficulty estimation, and recent work applies it to the agentic coding setting using LLM embeddings and rubric scores~\cite{ge2026agent}. 
However, their approach involves nondeterministic features that complicate reproducibility.

To address this gap, we conducted a large-scale empirical study on CoderForge-Preview~\cite{CoderForge2026} to answer the following research questions:
\begin{description}
  \item[\textbf{RQ1}] Can static task features predict agent success?
  \item[\textbf{RQ2}] Which feature groups account for predictive performance?
  \item[\textbf{RQ3}] Which individual features are most strongly associated with agent success?
\end{description}

We represent task difficulty as agent success rate and engineer several features across three groups: the gold patch, the repository, and the natural-language prompt, grounded in prior literature on code change complexity, software navigation, and requirements engineering. 
We then train models, measure their predictive accuracy, ablate feature groups, and decompose predictions using SHAP values to identify which features drive difficulty and how.
 
To this end, our work makes the following contributions:
\begin{itemize}
  \item We introduce a deterministic measurement framework for characterizing issue-resolution task difficulty from static patch, repository, and prompt properties.
  \item We provide large-scale evidence that agent success is predictable from static task structure.
  \item We identify factors that influence task success and briefly discuss how this varies across different levels of difficulty.
\end{itemize}

\section{Related Work}
\subsection{Agentic Benchmarks for Issue Resolution} 
 SWE-bench~\cite{jimenez2024swe} established the standard task for evaluating LLM-based agents: given a repository and an issue description, produce a patch that passes a held-out test suite. Progress on the benchmark has been rapid, but so have concerns about what aggregate solve rates actually measure. SWEBench+~\cite{aleithan2024swe} and UTBoost~\cite{yu2025utboost} independently found that a large fraction of resolved instances involved solution leakage or tests too weak to discriminate correct from incorrect patches; once filtered, resolution rates drop by more than half. SWE-bench Verified~\cite{openai2024verified} addressed this through manual developer vetting of 500 instances. Subsequent work has moved toward harder settings: long-horizon multi-file tasks in SWE-bench Pro~\cite{deng2025swe}, terminal-interaction tasks in Terminal Bench~\cite{merrill2026terminal}, and software performance optimization in GSO~\cite{shetty2026gso}. In parallel, training-oriented environments have scaled task supply: R2E-Gym~\cite{jain2025r2e}, SWE-Smith~\cite{yang2026swe}, and SWE-Rebench~\cite{badertdinov2026swe} together generate tens of thousands of executable tasks from real commits. CoderForgePreview~\cite{CoderForge2026}, the dataset we study, assembles 51K tasks from all three sources.
 What none of this work provides is an account of \emph{why} some tasks are harder than others. Our study is the first to systematically quantify which structural properties of a task derived from the required patch, the repository, and the issue description drive agent success rates on a large scale.

\subsection{Predicting Task Difficulty}
Item Response Theory (IRT) has become the standard tool for modeling
task-level difficulty in language-model evaluation~\cite{lalor2016building}, with recent work using it to reduce evaluation cost through adaptive task selection~\cite{polo2024tinybenchmarks,hofmann2025fluid,truong2025reliable}. Ge et al.~\cite{ge2026agent} bring IRT explicitly into the agentic coding setting, augmenting the Rasch model with LLM embeddings and LLM as-a-judge rubric scores to predict task difficulty on SWE-bench Verified, SWE-bench Pro, Terminal-Bench, and GSO.
 
Our features are fully deterministic measures grounded in empirical software engineering constructs rather than LLM encodings.
This makes our difficulty scores reproducible without any model inference. The structural features we identify could serve as cost effective, interpretable inputs to an IRT difficulty predictor of the type they describe.

\subsection{Code Change Complexity and Fault Localization}
\label{related:code}
The patch and repository features we engineer are grounded in a long empirical literature on what makes software changes hard to understand and fix. Munson and Elbaum~\cite{munson1998code} and Nagappan and
Ball~\cite{nagappan2005use} established code churn as a reliable predictor of defect density. The just-in-time defect prediction literature~\cite{kamei2016studying,mockus2000predicting,zhao2023systematic} further shows that it is not raw volume but \emph{diffusion}, how scattered a change is across files and hunks, that most consistently distinguishes defect-inducing from clean commits. We carry these measures into the agentic setting and find similar patterns.

\subsection{Linguistic Properties}
The natural-language prompt an agent receives is functionally a software requirement. Two decades of work has discovered ways to measure specification quality and clarity. They identified coordination
ambiguity, referential ambiguity, and attachment ambiguity as the three most consequential sources of misinterpretation in natural-language
requirements~\cite{handbook2003contract,ratnaparkhi1994maximum}.
Recent empirical work shows that LLMs struggle to reliably identify ambiguous requirements~\cite{larbi2025prompts} and that ambiguous specifications increase code generation uncertainty by introducing multiple valid interpretations~\cite{vijayvargiya2025interactive}.
What none of this work establishes is the \emph{relative} importance of specification quality compared to structural task properties, or whether linguistic features carry independent signal once structural features are controlled for. We directly address this in our work.

\section{Method}

\subsection{Dataset}
CoderForge-Preview~\cite{CoderForge2026} is the largest open dataset of coding agent trajectories to date, consisting of 258K test-verified trajectories spanning 51K tasks across 1,655 repositories. 
Tasks are drawn from three sources: R2E-Gym (4,216 tasks)~\cite{jain2025r2e}, SWE-Smith (37,221 tasks)~\cite{yang2026swe}, and SWE-Rebench (9,764 tasks)~\cite{badertdinov2026swe}. 
We generated trajectories using Qwen3-Coder-480B~\cite{yang2025qwen3}, one of the best-performing open-source models on agentic coding benchmarks, operating within an OpenHands v0.52.1 scaffold where the agent iteratively issues bash commands and file edits for up to 100 steps within an isolated Docker container. 
This makes CoderForge-Preview well-suited for studying the behavioral characteristics of capable coding agents rather than the failure modes of weaker models.
They generated multiple trajectories per task (up to 8), and retained those whose final patches pass all repository tests as successful. 
For our work, we excluded R2E-Gym tasks due to lack of a compatible patch format. We also excluded tasks for which we could not map to the repository using the GitHub API.
Our experiments therefore draw on 45,769 tasks from SWE-Smith and SWE-Rebench spanning 1,553 repositories, with a mean of 4.8 trajectories per task. 49.8\% of tasks are solved by every trajectory (\emph{always pass}), 32.5\% by none (\emph{always fail}), and 17.7\% exhibit mixed outcomes where some but not all trajectories succeed.

\subsection{Feature Engineering}
\label{sec:features}

 We engineered features across three groups, each capturing a distinct dimension of task difficulty.
\textbf{Patch features} characterize what a correct solution looks like.
\textbf{Repository features} characterize the structural environment the agent must navigate.
\textbf{Prompt features} characterize the natural-language specification the agent must interpret.
An overview of the three groups and their sub-constructs is shown in Table~\ref{tab:feature-overview}. Tables~\ref{tab:patch-features}--\ref{tab:prompt-features} list every feature with a description. 
Features marked $\dagger$ were dropped during the feature selection process (Section~\ref{sec:vif}); the 54 retained features are used in all models.

\begin{table}[!hbtp]
\centering
\caption{Feature taxonomy overview.
Each cell shows the sub-construct and its feature count}
\label{tab:feature-overview}
\footnotesize
\setlength{\tabcolsep}{4pt}
\begin{tabular}{lll}
\hline
\textbf{Patch} (18) &
\textbf{Repository} (14) &
\textbf{Prompt} (31) \\
\hline
Edit volume (3)           & Scale (5)                  & Component load (4) \\
Edit fragmentation (4)    & Directory structure (4)    & Coordinative complexity (6) \\
File scope (4)            & Naming ambiguity (1)       & Coordination ambiguity (5) \\
Churn distribution (3)    & Test infrastructure (3)    & Attachment ambiguity (7) \\
Spatial distribution (4)  & Documentation (1)          & Referential ambiguity (4) \\
                          &                            & Cohesion (5) \\
\hline
\end{tabular}
\end{table}
 
\subsubsection{Patch Features}

The patch group captures \emph{solution complexity}: how much the agent must find, understand, and change to resolve an issue.
The features here draw on the JIT defect prediction constructs discussed in Section~\ref{related:code}.
 
\begin{table}[!hbtp]
\centering
\caption{Patch features}
\label{tab:patch-features}
\footnotesize
\begin{tabular}{llp{5.2cm}}
\hline
\textbf{Sub-construct} & \textbf{Feature} & \textbf{What it measures} \\
\hline
\multirow{3}{*}{Edit volume}
  & \texttt{patch\_lines\_added}              & Lines inserted in the gold patch \\
  & \texttt{patch\_lines\_deleted}            & Lines removed in the gold patch \\
  & \texttt{patch\_lines\_changed\_total}$^\dagger$ & Total churn\\
\hline
\multirow{4}{*}{Edit fragmentation}
  & \texttt{patch\_num\_hunks}            & Total disjoint edit blocks \\
  & \texttt{patch\_edit\_fragmentation}   & Hunks per changed file \\
  & \texttt{patch\_hunk\_gap\_mean}        & Mean unchanged lines between consecutive hunks \\
  & \texttt{patch\_test\_files\_changed}  & Test files in the diff \\
\hline
\multirow{4}{*}{File scope}
  & \texttt{patch\_src\_files\_changed}        & Non-test source files changed \\
  & \texttt{patch\_python\_files\_changed}     & Python files changed \\
  & \texttt{patch\_extensions\_changed\_count} & Distinct file types touched \\
  & \texttt{patch\_num\_files\_changed}$^\dagger$ & Total files changed\\
\hline
\multirow{3}{*}{Churn distribution}
  & \texttt{patch\_churn\_gini\_over\_files}   & Edit concentration across files \\
  & \texttt{patch\_is\_single\_file}           & Binary: all edits in one file \\
  & \texttt{patch\_entropy\_over\_files}$^\dagger$ & Edit entropy \\
\hline
\multirow{4}{*}{Spatial distribution}
  & \texttt{patch\_primary\_file\_depth}          & Directory depth of most-edited file \\
  & \texttt{patch\_num\_unique\_dirs\_changed}    & Distinct directories touched \\
  & \texttt{patch\_mean\_file\_depth}$^\dagger$   & Mean depth of changed files \\
  & \texttt{patch\_max\_file\_depth}$^\dagger$    & Max depth of changed files \\
\hline
\end{tabular}
\end{table}
 
\subsubsection{Repository Features}
The repository group captures \emph{navigation difficulty}: how hard it is to locate the relevant code before making any edit. 
Fault localization is a prerequisite for patch generation; before an agent can write a fix, it must identify which files and lines are relevant to the issue. 
We represent navigation difficulty through two primary sub-constructs. 
\emph{Scale} features (file count, codebase size, directory count) reflect the size of the search space the agent must traverse. 
\emph{Directory structure} features (root-level breadth, maximum and mean nesting depth) reflect the structural complexity of that space. 
A \emph{naming ambiguity} feature (basenames shared by two or more files) captures another layer of complexity in identifying the right file to edit. 
\emph{Test infrastructure} and \emph{documentation} features capture repository organization conventions that may provide or withhold contextual signals useful for localization.

\begin{table}[!hbtp]
\centering
\caption{Repository features.}
\label{tab:repo-features}
\footnotesize
\begin{tabular}{llp{5.2cm}}
\hline
\textbf{Sub-construct} & \textbf{Feature} & \textbf{What it measures} \\
\hline
\multirow{5}{*}{Scale}
  & \texttt{repo\_file\_count}             & Total files in the repository \\
  & \texttt{repo\_python\_file\_count}     & Python files \\
  & \texttt{repo\_total\_known\_size}      & Codebase size in bytes \\
  & \texttt{repo\_mean\_known\_file\_size} & Mean file size in bytes \\
  & \texttt{repo\_dir\_count}              & Total directory count \\
\hline
\multirow{4}{*}{Directory structure}
  & \texttt{repo\_top\_level\_dir\_count} & Breadth at root level \\
  & \texttt{repo\_max\_depth}             & Maximum directory nesting depth \\
  & \texttt{repo\_mean\_file\_depth}      & Mean file nesting depth \\
  & \texttt{repo\_test\_ratio}            & Fraction of files that are tests \\
\hline
Naming ambiguity
  & \texttt{repo\_basename\_collision\_count} & Basenames shared by $\geq$2 files \\
\hline
\multirow{3}{*}{Test infrastructure}
  & \texttt{repo\_test\_file\_count}   & Number of test files \\
  & \texttt{repo\_has\_tests\_dir}     & Dedicated \texttt{tests/} directory present \\
  & \texttt{repo\_has\_examples\_dir}  & \texttt{examples/} directory present \\
\hline
Documentation
  & \texttt{repo\_has\_docs\_dir} & \texttt{docs/} directory present \\
\hline
\end{tabular}
\end{table}
 
\subsubsection{Prompt Features}
The prompt group captures \emph{specification interpretability}: whether the agent can derive an action plan from the issue text. 
We draw on the RE literature's taxonomy of natural-language defects to motivate four sub-constructs.
\emph{Component load} features measure how much information the agent must integrate.
\emph{Coordinative complexity} features represent structural processing difficulty at the sentence level, motivated by Gibson's Dependency Locality Theory~\cite{gibson2000dependency} and the finding that syntactically complex prompts increase LLM code generation
uncertainty~\cite{larbi2025prompts,vijayvargiya2025interactive}.
\emph{Ambiguity} features represent the three defect types identified by the RE literature as most consequential for misinterpretation: coordination ambiguity (scope of conjunctions), attachment ambiguity (which constituent a modifier modifies), and referential ambiguity (which noun phrase a pronoun resolves to)~\cite{handbook2003contract,yang2010methodology}.
\emph{Cohesion} features measure lexical and semantic connectivity between adjacent sentences~\cite{Halliday2014-el,mcnamara1996good}.
 
\begin{table}[!hbtp]
\centering
\caption{Prompt features.}
\label{tab:prompt-features}
\footnotesize
\begin{tabular}{llp{3cm}}
\hline
\textbf{Sub-construct} & \textbf{Feature} & \textbf{What it measures} \\
\hline
\multirow{4}{*}{Component load}
  & \texttt{prompt\_len\_bpe}                         & Prompt length in BPE tokens \\
  & \texttt{prompt\_sent\_count}                      & Number of sentences \\
  & \texttt{prompt\_clauses\_per\_sentence}$^\dagger$ & Clauses per sentence \\
  & \texttt{prompt\_content\_tokens\_per\_sentence}$^\dagger$ & Content tokens per sentence \\
\hline
\multirow{6}{*}{Coordinative complexity}
  & \texttt{prompt\_dep\_distance\_mean}          & Mean linear dependency distance \\
  & \texttt{prompt\_dep\_distance\_max}           & Max linear dependency distance \\
  & \texttt{prompt\_branching\_factor\_mean}      & Mean syntactic branching factor \\
  & \texttt{prompt\_clause\_nesting\_depth\_mean} & Mean clause nesting depth \\
  & \texttt{prompt\_clause\_length\_mean}         & Mean clause length \\
  & \texttt{prompt\_clausal\_branching\_factor}   & Mean clause-level branching \\
\hline
\multirow{5}{*}{Coordination ambiguity}
  & \texttt{prompt\_coordination\_density}     & Conjunctions per clause \\
  & \texttt{prompt\_mean\_conj\_chain\_len}    & Mean conjunction chain length \\
  & \texttt{prompt\_max\_conj\_chain\_len}     & Max conjunction chain length \\
  & \texttt{prompt\_verb\_coordination\_ratio} & Conjunctions with verb heads \\
  & \texttt{prompt\_noun\_coordination\_ratio} & Conjunctions with noun heads \\
\hline
\multirow{7}{*}{Attachment ambiguity}
  & \texttt{prompt\_pp\_density\_per\_noun}              & PP count per noun \\
  & \texttt{prompt\_pp\_density\_per\_clause}            & PP count per clause \\
  & \texttt{prompt\_mean\_pp\_chain\_depth}              & Mean PP nesting depth \\
  & \texttt{prompt\_max\_pp\_chain\_depth}               & Max PP nesting depth \\
  & \texttt{prompt\_multi\_pp\_head\_ratio}              & Heads with $\geq$2 attached PPs \\
  & \texttt{prompt\_attachment\_competition\_ratio}      & Heads with $\geq$2 competing dependents \\
  & \texttt{prompt\_mean\_competing\_dependents\_per\_head} & Mean competing dependents per head \\
\hline
\multirow{4}{*}{Referential ambiguity}
  & \texttt{prompt\_pronouns\_per\_sentence}                & Pronouns per sentence \\
  & \texttt{prompt\_pronoun\_candidate\_antecedent\_density} & Candidate antecedents per pronoun \\
  & \texttt{prompt\_multi\_antecedent\_pronoun\_ratio}      & Pronouns with $\geq$2 antecedents \\
  & \texttt{prompt\_pronoun\_ratio}$^\dagger$               & Pronoun fraction \\
\hline
\multirow{5}{*}{Cohesion}
  & \texttt{prompt\_adjacent\_lemma\_overlap\_mean}  & Lemma overlap between adjacent sentences \\
  & \texttt{prompt\_sentence\_embedding\_sim\_mean}  & Mean cosine sim of adjacent sentences \\
  & \texttt{prompt\_sentence\_embedding\_sim\_min}   & Min cosine sim of adjacent sentences \\
  & \texttt{prompt\_entity\_reuse\_ratio}            & Entity reuse across sentences \\
  & \texttt{prompt\_coref\_chain\_len\_mean}$^\dagger$ & Coreference chain length \\
\hline
\end{tabular}
\end{table}

\subsection{Feature Selection}
\label{sec:vif}
We applied iterative Variance Inflation Factor (VIF)~\cite{belsley2005regression} filtering with threshold 10~\cite{marquardt1970generalized}, dropping the highest-VIF feature in each iteration until all scores fall below the threshold.
This process dropped 9 features (marked $\dagger$ in Tables~\ref{tab:patch-features}--\ref{tab:prompt-features}), leaving 54 features with a maximum VIF of 7.59.
The same feature set is used for all three models.

\subsection{Predictive Modeling}
\subsubsection{Outcome Variables.}
For each task, we compute three outcome variables: \textbf{(1) pass\_rate}: the fraction of runs that passed all tests (continuous, $[0,1]$; mean $= 0.593$, SD $= 0.455$).
\textbf{(2) any\_success} (\textit{pass@k}): an indicator of whether at least one run succeeded (positive rate: $67.5\%$).
\textbf{(3) maj\_success} (majority success): an indicator of whether the majority ($\geq 50\%$) of runs succeeded (positive rate: $61.7\%$).

\begin{table}[!hbtp]
\centering
\caption{Outcome variable distributions ($n = 45{,}7695$ tasks).}
\label{tab:outcome-dist}
\begin{subtable}[t]{0.40\linewidth}
    \centering
    \caption{Continuous outcome}
    \label{tab:outcome-continuous}
    \begin{tabular}{lcc}
    \hline
    \textbf{Outcome} & \textbf{Mean} & \textbf{SD} \\
    \hline
    \texttt{pass\_rate} & 0.593 & 0.455 \\
    \hline
    \end{tabular}
\end{subtable}
\begin{subtable}[t]{0.45\linewidth}
    \centering
    \caption{Binary outcomes}
    \label{tab:outcome-binary}
    \begin{tabular}{lcc}
    \hline
    \textbf{Outcome} & \textbf{Positive rate} & \textbf{Imbalance ratio} \\
    \hline
    \texttt{any\_success} & 67.5\% & 2.08 \\
    \texttt{maj\_success} & 61.7\% & 1.61 \\
    \hline
    \end{tabular}
\end{subtable}
\end{table}

\subsubsection{Data Split.}
We split the dataset at the task level into training and held-out test sets using an $80/20$ split, resulting in $36{,}615$ training tasks and $9{,}154$ test tasks.

\subsubsection{Model.}
We trained and three model classes for each outcome. We selected hyperparameters using  \texttt{RandomizedSearchCV} with 4-fold inner cross-validation and 28 iterations per model group, using ROC-AUC as the classification scoring criterion and negative mean squared error for regression. 
All hyperparameters reported below reflect the tuned values found by this procedure.
 
\begin{itemize}
\item \textbf{Binary classification models.}
For \texttt{any\_success} and \texttt{maj\_success}, we train: (i) Logistic Regression with L2 regularization and standardized features ($C = 0.013$ for \texttt{any\_success}; $C = 0.037$ for \texttt{maj\_success}); (ii) Random Forest with 500 trees, maximum depth 16, \texttt{max\_features}$=0.25$, and minimum 8 samples per leaf; and (iii) XGBoost with 400 trees, maximum depth 8, learning rate 0.03, 
and minimum child weight 2. 
Ensemble methods, particularly Random Forest and gradient boosting, consistently outperform linear classifiers in software defect prediction tasks~\cite{lessmann2008benchmarking,ghotra2015revisiting}, motivating their inclusion alongside the interpretable linear baseline.
 
\item \textbf{Regression models.}
For \texttt{pass\_rate}, we train: (i) Ridge Regression with $\alpha = 115.67$ and standardized features; (ii) Random Forest Regressor with 500 trees, unlimited depth, \texttt{max\_features}$=0.25$, and minimum 4 samples per leaf; and (iii) XGBoost Regressor with a squared-error objective and the same hyperparameters as the classifiers.
\end{itemize}

\subsubsection{Evaluation}
\noindent\textbf{Output Metrics.} We evaluated the models on the held-out test set. 
Classification performance is reported using AUC--ROC, Precision--Recall AUC (PR--AUC), Brier score, F1, and Matthews Correlation Coefficient (MCC). 
Regression performance is reported using RMSE, MAE, and $R^2$. 
We additionally perform 10-fold cross-validation on the training set for the XGBoost model and report mean $\pm$ SD to confirm stability.\\

\noindent\textbf{Feature Importance.}
Beyond predictive accuracy, we quantify the direction and magnitude of each feature's association with task success.

\begin{itemize}

\item \textbf{Component ablation.}
We trained XGBoost models using each feature group alone, all pairwise combinations, and the full three-way combination. 
We report test AUC and $R^2$ for each subset to estimate how much predictive signal is contributed by patch complexity, repository structure, and issue text.

\item \textbf{SHAP feature importance.}
We compute SHAP values for the best-performing XGBoost model and rank features by their mean absolute contribution to model predictions. 
SHAP additive explanations provide a theoretically grounded decomposition of model predictions~\cite{lundberg2017unified}, making them suitable for interpreting the relative importance of features in a non-linear ensemble model.

\end{itemize}

\section{Results}
\subsection{Task-Level Feature Agent Success Prediction Accuracy (RQ1)}
 
Static task features derived from the gold patch, repository structure, and agent prompt yield strong predictive performance across all three outcome targets. 
Table~\ref{tab:heldout-classification} reports held-out test performance for all models alongside a majority-class baseline that assigns the positive label to every task. 
The baseline achieves AUC $= 0.500$ and MCC $= 0.000$, confirming that it carries no discriminative information; its comparatively high F1 score of 0.806 is due to the 67.5\% positive rate in the \texttt{any\_success} target and should not be interpreted as predictive signal.
 
Among the three classifiers, XGBoost and Random Forest reach near-identical performance on \texttt{any\_success} (XGBoost: AUC $= 0.863$, MCC $= 0.549$, Brier $= 0.129$; Random Forest: AUC $= 0.863$, MCC $= 0.538$, Brier $= 0.128$).
 
\begin{table}[!hbtp]
\centering
\caption{Held-out test performance. Majority denotes the always-positive baseline.
Bold denotes best non-baseline result per metric per target.}
\label{tab:heldout-classification}
\begin{tabular}{llccccc}
\hline
\textbf{Target} & \textbf{Model} & \textbf{AUC} & \textbf{PR-AUC} & \textbf{Brier} & \textbf{F1} & \textbf{MCC} \\
\hline
\texttt{any\_success} & Majority       & 0.500 & 0.675 & 0.219 & 0.806 & 0.000 \\
 & Logistic       & 0.750 & 0.849 & 0.183 & 0.741 & 0.341 \\
 & Random Forest  & \textbf{0.863} & 0.914 & \textbf{0.128} & 0.832 & 0.538 \\
 & XGBoost        & \textbf{0.863} & \textbf{0.916} & 0.129 & \textbf{0.848} & \textbf{0.549} \\
\hline
\texttt{maj\_success} & Majority       & 0.500 & 0.617 & 0.383 & 0.763 & 0.000 \\
 & Logistic       & 0.738 & 0.809 & 0.198 & 0.722 & 0.330 \\
 & Random Forest  & \textbf{0.848} & \textbf{0.885} & \textbf{0.147} & 0.797 & 0.508 \\
 & XGBoost        & 0.845 & 0.884 & 0.148 & \textbf{0.811} & \textbf{0.517} \\
\hline
\end{tabular}
 
\medskip
\begin{tabular}{lccc}
\hline
\multicolumn{4}{l}{\textit{\texttt{pass\_rate} regression}} \\
\textbf{Model} & \textbf{RMSE} & \textbf{MAE} & $\boldsymbol{R^2}$ \\
\hline
Ridge          & 0.415 & 0.374 & 0.169 \\
Random Forest  & \textbf{0.350} & \textbf{0.276} & \textbf{0.409} \\
XGBoost        & \textbf{0.350} & \textbf{0.276} & 0.408 \\
\hline
\end{tabular}
\end{table}

The substantially lower performance of Logistic Regression (AUC $= 0.750$, MCC $= 0.341$) indicates that the relationship between task features and agent success is \textit{non-linear} and cannot be adequately captured by a linear decision boundary.
 
Predicting the continuous \texttt{pass\_rate} outcome is a harder problem. 
Ridge regression accounts for only 16.9\% of variance ($R^2 = 0.169$), whereas Random Forest and XGBoost reach $R^2 = 0.409$ and $R^2 = 0.408$ respectively, showing a $2.4\times$ improvement over the linear baseline. 
The two ensemble methods are essentially tied on this target. 
A portion of the remaining variance is likely irreducible: 17.7\% of tasks show mixed outcomes across repeated runs, and no static task-level feature can account for the stochastic component of agent behavior within a fixed task.

Ten-fold cross-validation on the XGBoost training set confirms that held-out test performance is stable. 
For \texttt{any\_success}, mean CV AUC is $0.851 \pm 0.009$ and mean CV MCC is $0.558 \pm 0.017$, both in 
close agreement with held-out test values. For \texttt{maj\_success}, mean CV AUC  is $0.838 \pm 0.006$ and mean CV MCC is $0.530 \pm 0.008$. For the \texttt{pass\_rate} regression, mean CV $R^2$ is $0.386 \pm 0.008$, consistent 
with the held-out result of $0.408$.

\begin{figure}[!hbtp]
  \centering
  \includegraphics[width=0.5\linewidth]{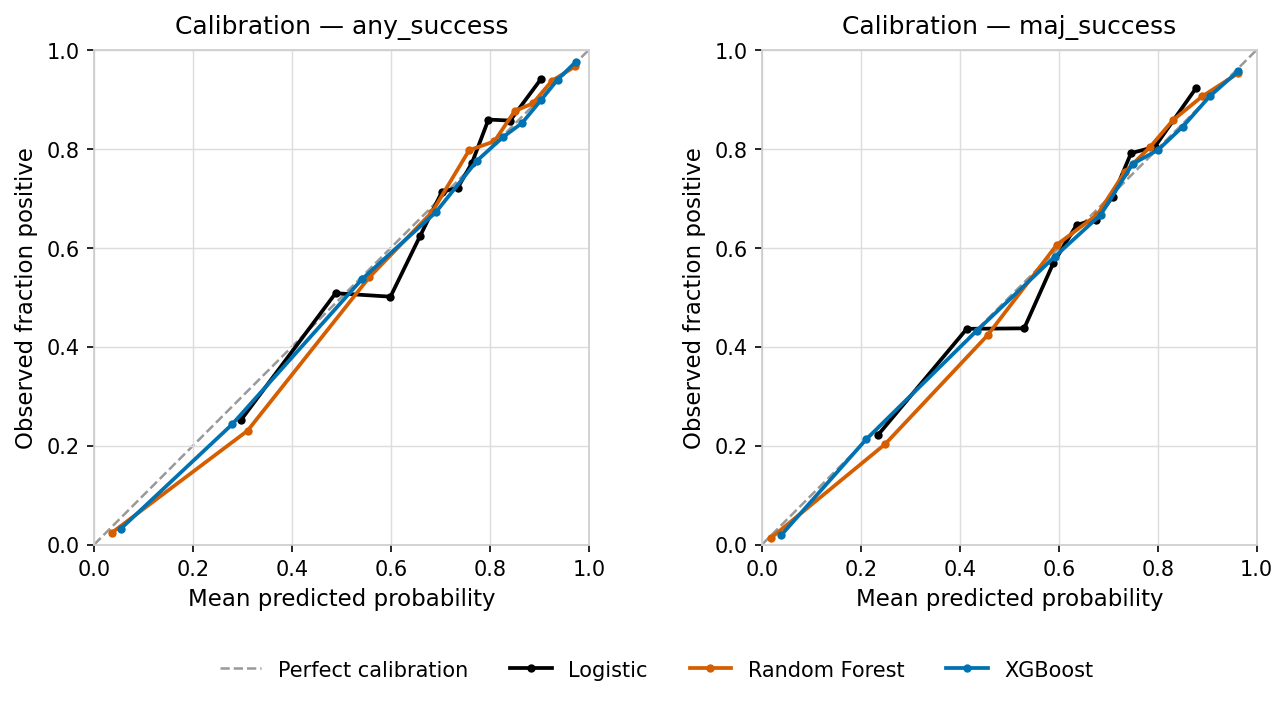}
  \caption{Calibration reliability diagrams}
  \label{fig:calibration-both}
\end{figure}

Calibration reliability diagrams for all three classifiers are shown in Figure~\ref{fig:calibration-both}. 
XGBoost is the best-calibrated model across both outcomes, with small deviations from the diagonal.
Random Forest is well-calibrated in the upper probability range but exhibits overconfidence in the low-to-mid range, where predicted probabilities consistently exceed observed success rates. 
Logistic Regression shows the largest deviations across both
outcomes, consistent with its compressed prediction range and weaker discriminative performance. 
The well-calibrated XGBoost probability estimates make it the preferred model for downstream applications such as task stratification and difficulty-weighted evaluation.

\begin{tcolorbox}[
  colback=lipicsLightGray,
  colframe=lipicsLightGray,
  boxrule=0pt,
  arc=2pt,
  left=6pt, right=6pt, top=4pt, bottom=4pt
]
\textbf{Finding 1.}
Task-level features derived from the patch, repository, and agent prompt predict agent success with AUC $= 0.863$ and MCC $= 0.549$ (XGBoost), explaining 41\% of pass-rate variance ($R^2 = 0.408$), with well-calibrated probability estimates across the full prediction range.
\end{tcolorbox}

\subsection{Feature Groups Driving Predictive Performance (RQ2)}

To determine the independent contribution of each feature group, we trained XGBoost models using each group alone and in all pairwise and three-way combinations, evaluating on the held-out test set across all three outcome definitions. Table~\ref{tab:component-ablation} presents the results.

\begin{table*}[t]
  \centering
  \caption{Component ablation using XGBoost on the held-out test set.
           AUC and Brier score are reported for the two binary outcomes;
           $R^2$ and MAE for the continuous \texttt{pass\_rate} outcome.
           Bold denotes the best value in each column.}
  \label{tab:component-ablation}
  \begin{tabular}{lc cc cc cc}
    \hline
    & &
      \multicolumn{2}{c}{\texttt{any\_success}} &
      \multicolumn{2}{c}{\texttt{maj\_success}} &
      \multicolumn{2}{c}{\texttt{pass\_rate}} \\
    \hline
    \textbf{Feature set} & \textbf{N} &
      \textbf{AUC} & \textbf{Brier} &
      \textbf{AUC} & \textbf{Brier} &
      $\boldsymbol{R^2}$ & \textbf{MAE} \\
    \hline
    \hline
    Patch only        & 13 & 0.846 & 0.135 & 0.830 & 0.154 & 0.375 & 0.285 \\
    Repo only         & 14 & 0.839 & 0.137 & 0.820 & 0.158 & 0.350 & 0.291 \\
    Prompt only       & 27 & 0.599 & 0.214 & 0.593 & 0.231 & 0.025 & 0.417 \\
    \hline
    Patch + Repo      & 27 & 0.861 & 0.129 & 0.841 & 0.149 & 0.401 & 0.275 \\
    Patch + Prompt    & 40 & 0.850 & 0.135 & 0.832 & 0.154 & 0.380 & 0.290 \\
    Repo + Prompt     & 41 & 0.839 & 0.138 & 0.823 & 0.158 & 0.361 & 0.295 \\
    \hline
    All features      & 54 & \textbf{0.863} & \textbf{0.129} & \textbf{0.846} & \textbf{0.148} & \textbf{0.406} & \textbf{0.277} \\
    \hline
  \end{tabular}
\end{table*}

Patch features alone achieve AUC $= 0.846$ on \texttt{any\_success} and $R^2 = 0.375$ on \texttt{pass\_rate}. Repository features alone perform comparably, with AUC $= 0.839$ and $R^2 = 0.350$. Combining these two groups yields AUC $= 0.861$ and $R^2 = 0.401$, within 0.002 AUC and 0.006 $R^2$ of the full 54-feature model (AUC $= 0.863$, $R^2 = 0.406$).
 
Considered in isolation, the prompt group performs near chance on both binary outcomes (AUC $= 0.599$ for \texttt{any\_success}; AUC $= 0.593$ for \texttt{maj\_success}) and explains almost none of the continuous variance ($R^2 = 0.025$). 
Adding prompt features to patch alone yields marginal gains ($\Delta\mathrm{AUC} \leq 0.004$, $\Delta R^2 = 0.005$ on \texttt{pass\_rate}), and adding them to repository alone is similarly modest ($\Delta\mathrm{AUC} \leq 0.004$, $\Delta R^2 = 0.012$ on \texttt{pass\_rate}). 
Critically, once both patch and repository features are included, adding all prompt features produces negligible further gain: $\Delta\mathrm{AUC} \leq 0.002$ and $\Delta R^2 < 0.01$ across all three outcome definitions. 
This pattern is consistent across both discrimination (AUC) and calibration (Brier score) measures.
 
These results indicate that the difficulty of a task, as measured by agent success rates, is determined almost entirely by the structural properties of the required solution. Specifically, how complex and scattered the patch is, and how large and complex the repository environment is are the strongest indicators. 
The linguistic and syntactic properties of the agent prompt, as operationalized by our feature set, carry negligible independent signal beyond these structural factors. 
Whether prompt properties matter conditionally, for specific tasks or in specific circumstances, is a question this ablation study alone cannot answer; we examine individual feature contributions in RQ3.
 
\begin{tcolorbox}[
  colback=lipicsLightGray,
  colframe=lipicsLightGray,
  boxrule=0pt,
  arc=2pt,
  left=6pt, right=6pt, top=4pt, bottom=4pt
]
  \textbf{Finding~2.}
  Patch complexity and repository structure together account for virtually all of the predictable variance in agent success across all outcome definitions, achieving AUC $= 0.861$ and $R^2 = 0.401$ with 27 features -- within $0.002$ AUC of the full model.
\end{tcolorbox}

\subsection{Individual Feature Association with Agent Success (RQ3)}\label{sec:rq3}

We use SHAP values computed via \texttt{TreeExplainer} on the held-out test set to identify which features drive predictions and in what direction~\cite{lundberg2017unified}. We report mean absolute SHAP value as the importance measure, SHAP dependence plots for directional effects, and SHAP interaction values for joint effects between feature pairs. All analyses use the tuned XGBoost model; we compare feature rankings between \texttt{any\_success} and \texttt{pass\_rate} to assess whether findings generalize across outcome definitions.

\subsubsection{Feature Importance and Directional Effects}

\begin{figure}[!hbtp]
  \centering
  \includegraphics[width=\textwidth,keepaspectratio]{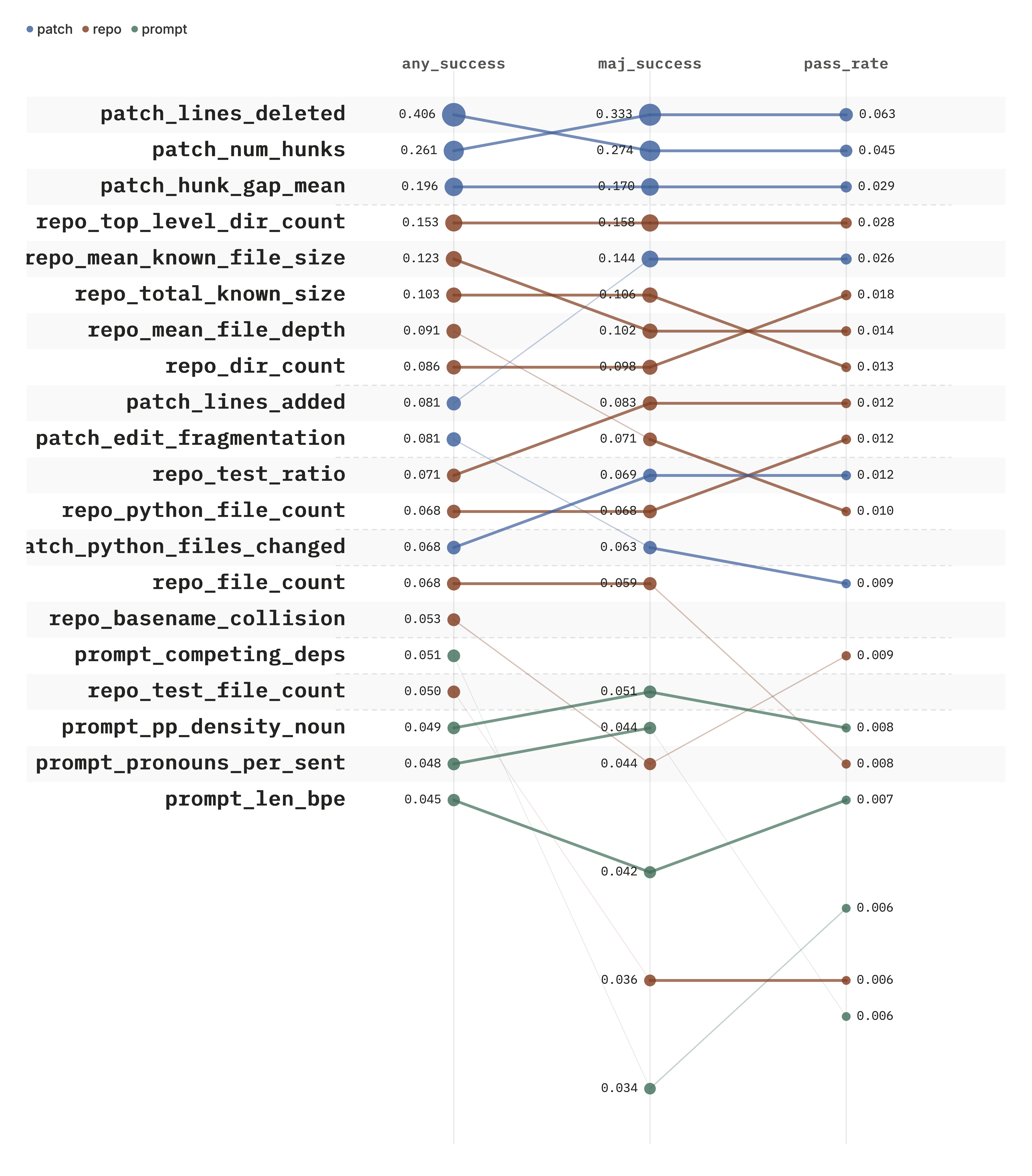}
  \caption{Rank--magnitude bump chart for the top 20 features by mean absolute SHAP value across three outcome targets (XGBoost, held-out test set). Each dot is one feature--outcome pair; dot area is proportional to mean absolute SHAP. Connecting lines trace rank shifts across outcomes; steep slopes indicate instability. Color encodes feature family: patch (blue), repo (red-brown), prompt (green). The top six ranks are stable across all three outcomes (near-flat lines); prompt features first appear at rank 16 with visibly smaller dots and show greater rank instability below rank 16.}
  \label{fig:shap-bump}
\end{figure}

\begin{figure}[!hbtp]
  \centering
  \includegraphics[height=\textheight]{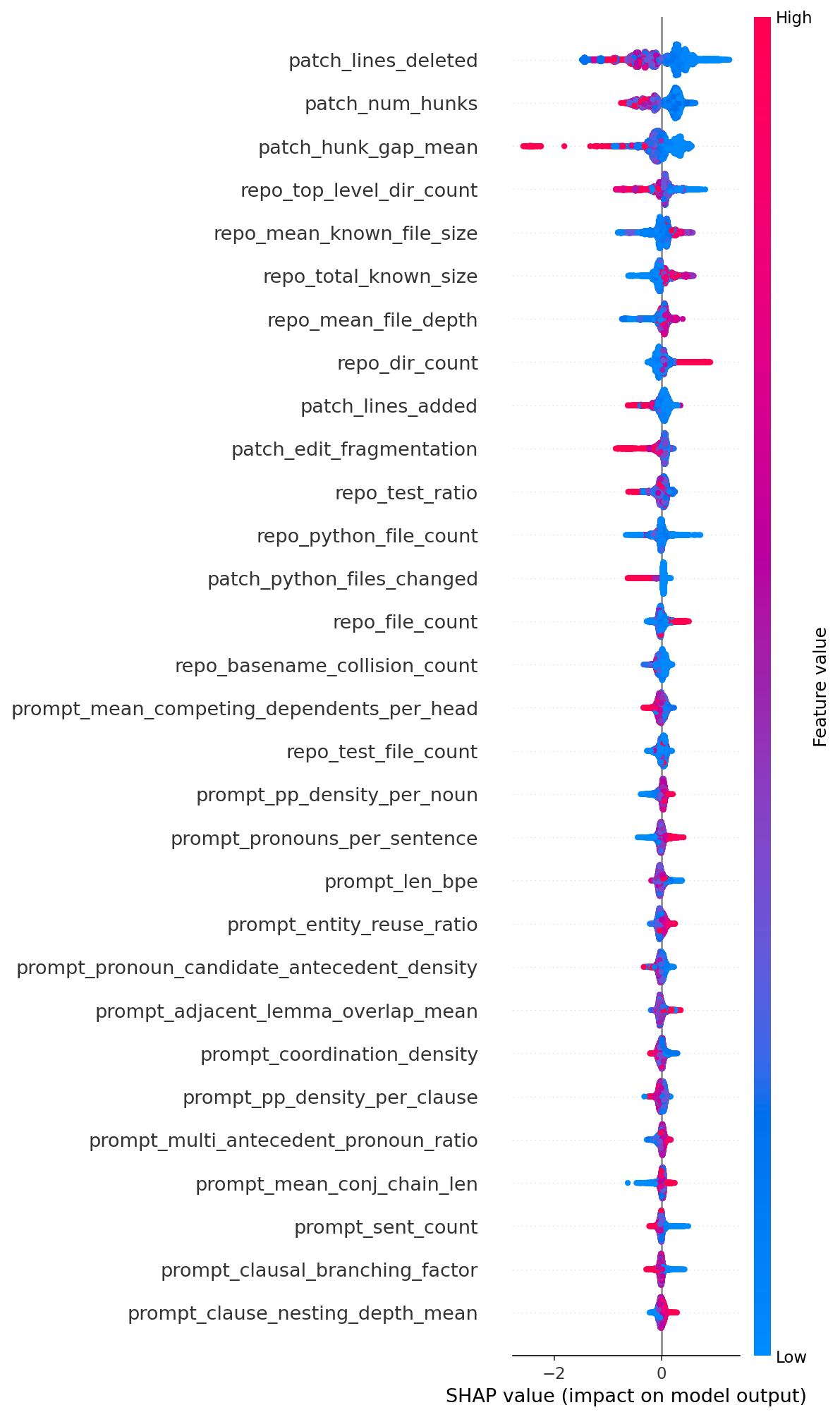}
  \caption{SHAP summary plot for \texttt{any\_success}.}
  \label{fig:shap-summary}
\end{figure}

Figure~\ref{fig:shap-bump} shows the top 20 features by mean absolute SHAP value for each outcome. 
Figure~\ref{fig:shap-summary} shows the corresponding SHAP summary plot for \texttt{any\_success}, which reveals the direction of each effect.
The same three features occupy the top positions across all outcomes: \texttt{patch\_lines\_deleted}, \texttt{patch\_num\_hunks}, and \texttt{patch\_hunk\_gap\_mean} (ranks 1--3, with only a 1/2 swap between targets, visible as near-flat connecting lines in Figure~\ref{fig:shap-bump}). 
For \texttt{any\_success}, \texttt{patch\_lines\_deleted} has a mean absolute SHAP of $0.406$-- $2.65\times$ larger than the fourth-ranked feature\\ (\texttt{repo\_top\_level\_dir\_count}, $0.153$)---and the top three features together account for 29\% of total mean absolute SHAP across all 54 features. 

In the summary, all three plots show a clear monotone negative pattern: high feature values push predictions toward failure.
\texttt{patch\_hunk\_gap\_mean} additionally exhibits a small number of extreme outliers (SHAP $\approx -2.5$), indicating that tasks with unusually large inter-hunk distances are especially penalized. 
SHAP values decay steeply below the leading group: the 20th-ranked feature (\texttt{prompt\_len\_bpe}) has a mean absolute SHAP $9\times$ smaller than the top feature, and below rank 10 distributions are tightly clustered near zero with no directional gradient. 
Ranks 4--14 are occupied entirely by repository structural and patch features, all showing negative directional effects in Figure~\ref{fig:shap-summary}. 
These contribute independently of the top patch trio, suggesting that the navigational complexity of the repository the agent must traverse matters beyond the properties of the edit itself. 

\begin{figure}[!hbtp]
  \centering
  \begin{subfigure}[t]{0.48\linewidth}
    \centering
    \includegraphics[width=\linewidth]{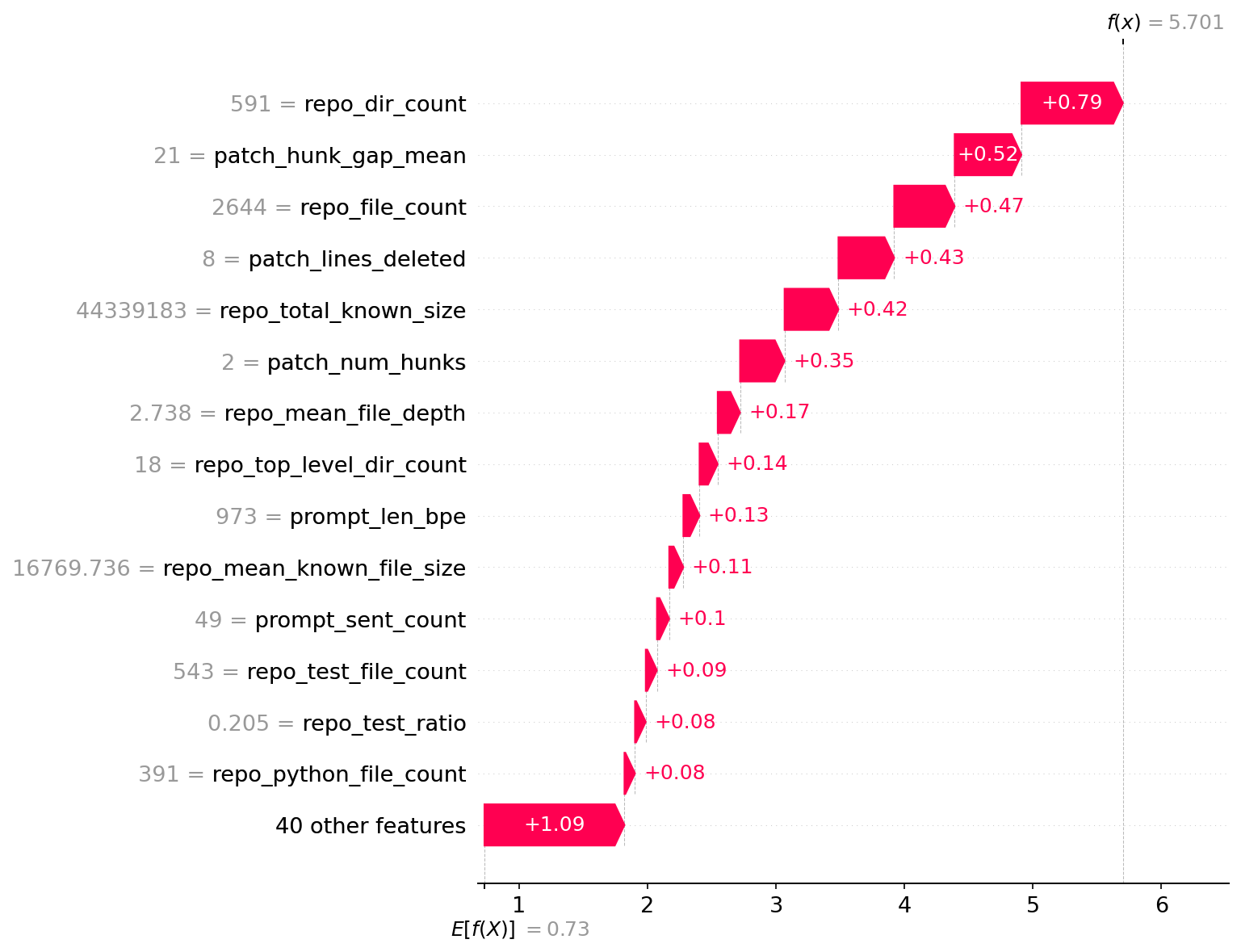}
    \caption{Easiest task (\texttt{pygments/pygments}); $\hat{p} = 0.997$,
      $E[f(x)] = 0.73$, observed: success. Every structural contributor pushes
      the prediction upward from baseline.}
    \label{fig:waterfall-easiest}
  \end{subfigure}
  \hfill
  \begin{subfigure}[t]{0.48\linewidth}
    \centering
    \includegraphics[width=\linewidth]{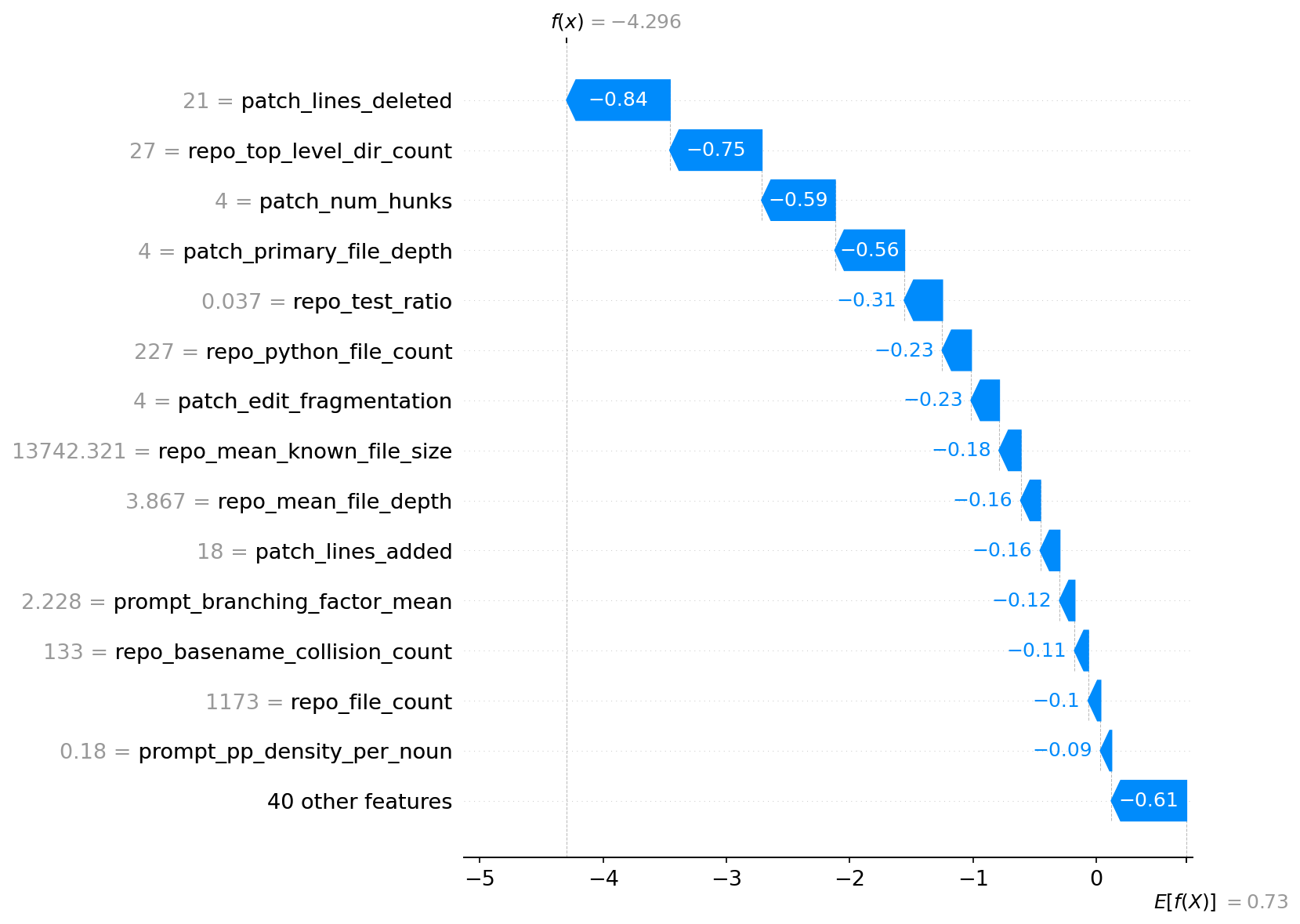}
    \caption{Hardest task (\texttt{getnikola/nikola}); $\hat{p} = 0.013$,
      $E[f(x)] = 0.73$, observed: failure. Despite only moderate patch volume,
      high \texttt{repo\_top\_level\_dir\_count} is the second-largest negative
      contributor, illustrating the patch--repository interaction.}
    \label{fig:waterfall-hardest}
  \end{subfigure}
  \caption{SHAP waterfalls for extreme-difficulty tasks (XGBoost,
    \texttt{any\_success} target). All values in log-odds space; $E[f(x)] = 0.73$ for both.}
  \label{fig:waterfalls-extreme}
\end{figure}

\begin{figure}[!hbtp]
  \centering
  \begin{subfigure}[t]{0.48\linewidth}
    \centering
    \includegraphics[width=\linewidth]{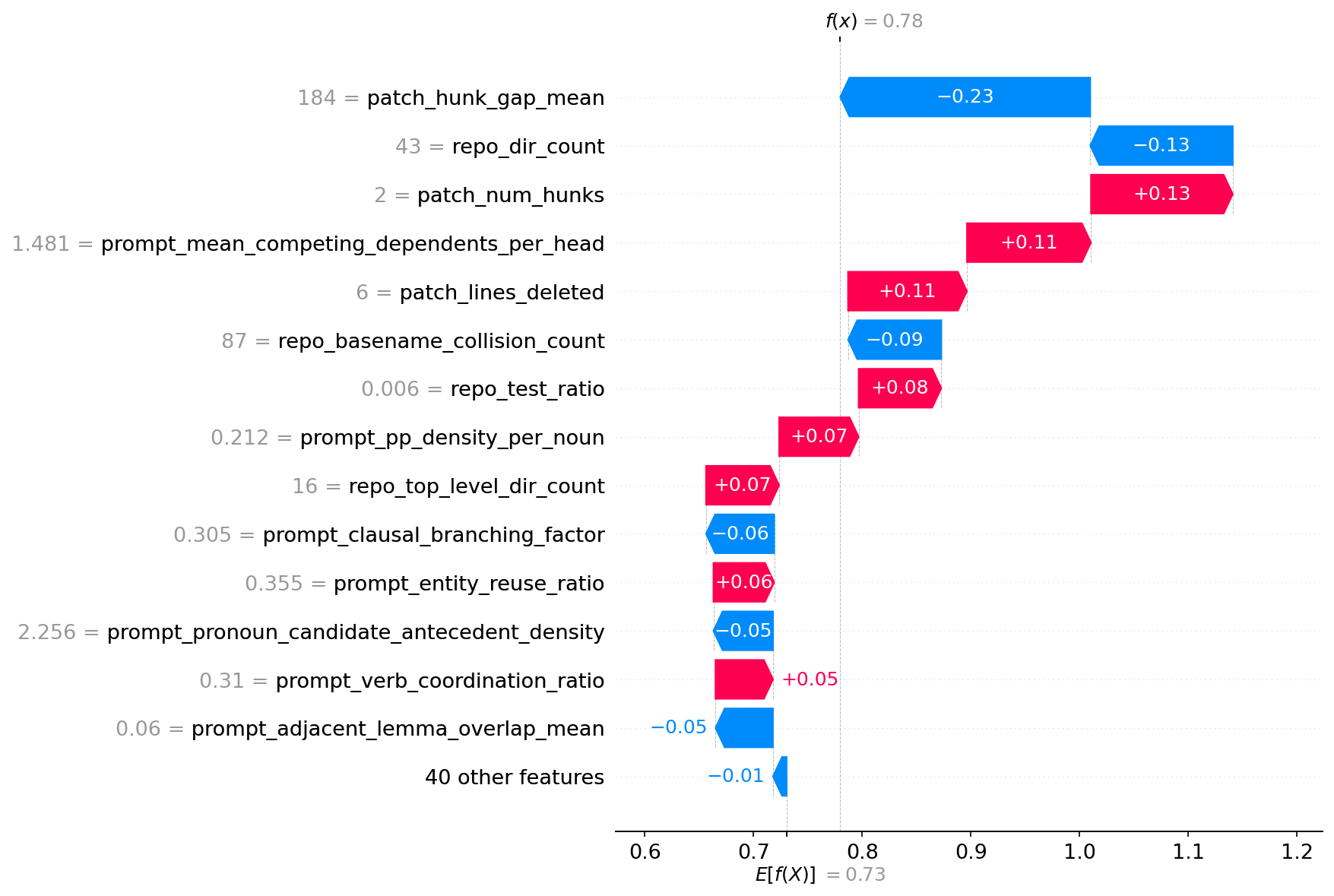}
    \caption{\texttt{kurtmckee/feedparser}; $\hat{p} = 0.684$, observed: success.
      Mixed structural signals sit just above baseline.}
    \label{fig:waterfall-average}
  \end{subfigure}
  \hfill
  \begin{subfigure}[t]{0.48\linewidth}
    \centering
    \includegraphics[width=\linewidth]{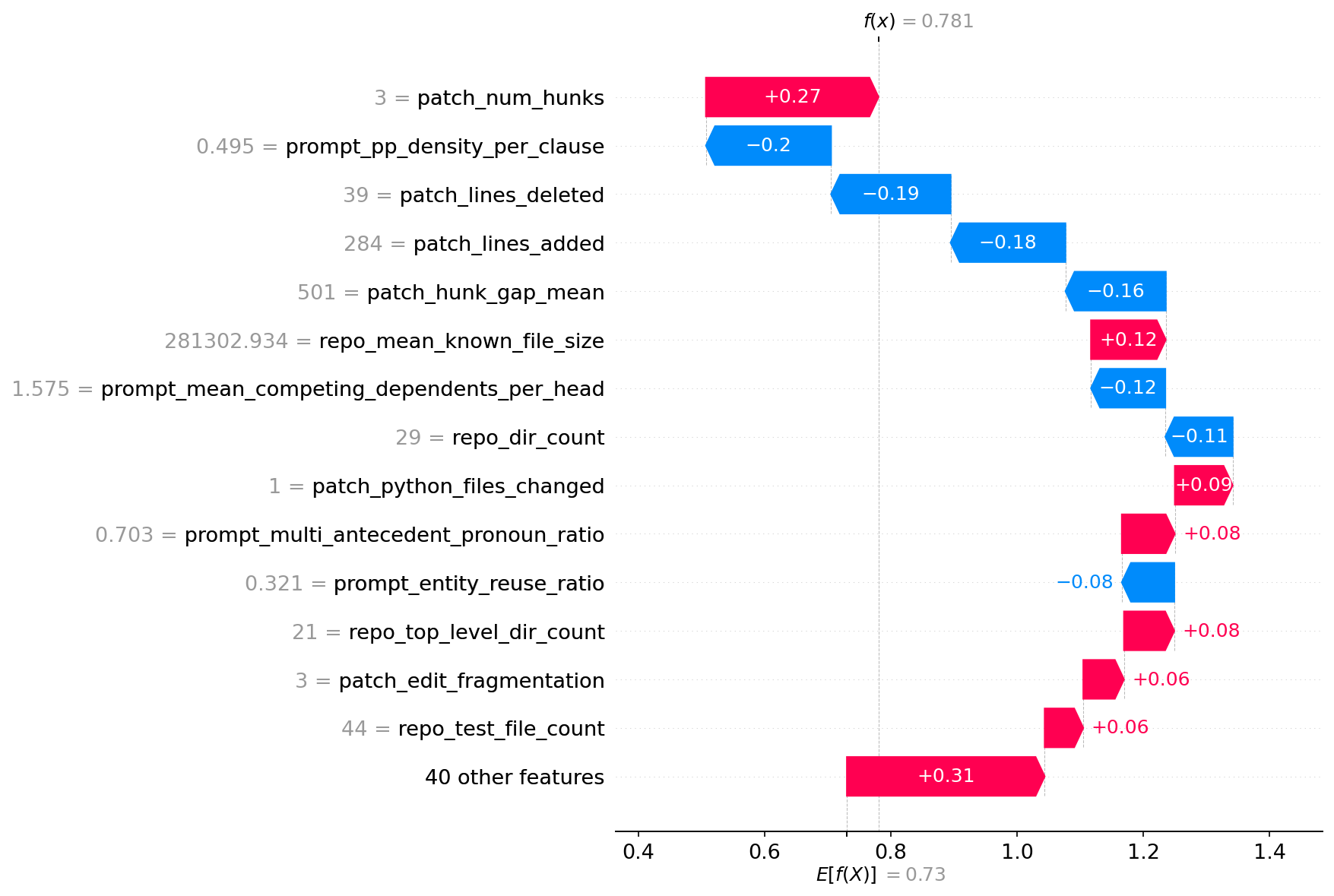}
    \caption{\texttt{sqlglot}; $\hat{p} = 0.685$, observed: failure.
      \texttt{prompt\_pp\_density\_per\_clause} ($-0.20$) is the second-largest
      contributor.}
    \label{fig:waterfall-avg-sqlglot}
  \end{subfigure}
  \vspace{-0.3em}
  \begin{subfigure}[t]{0.48\linewidth}
    \centering
    \includegraphics[width=\linewidth]{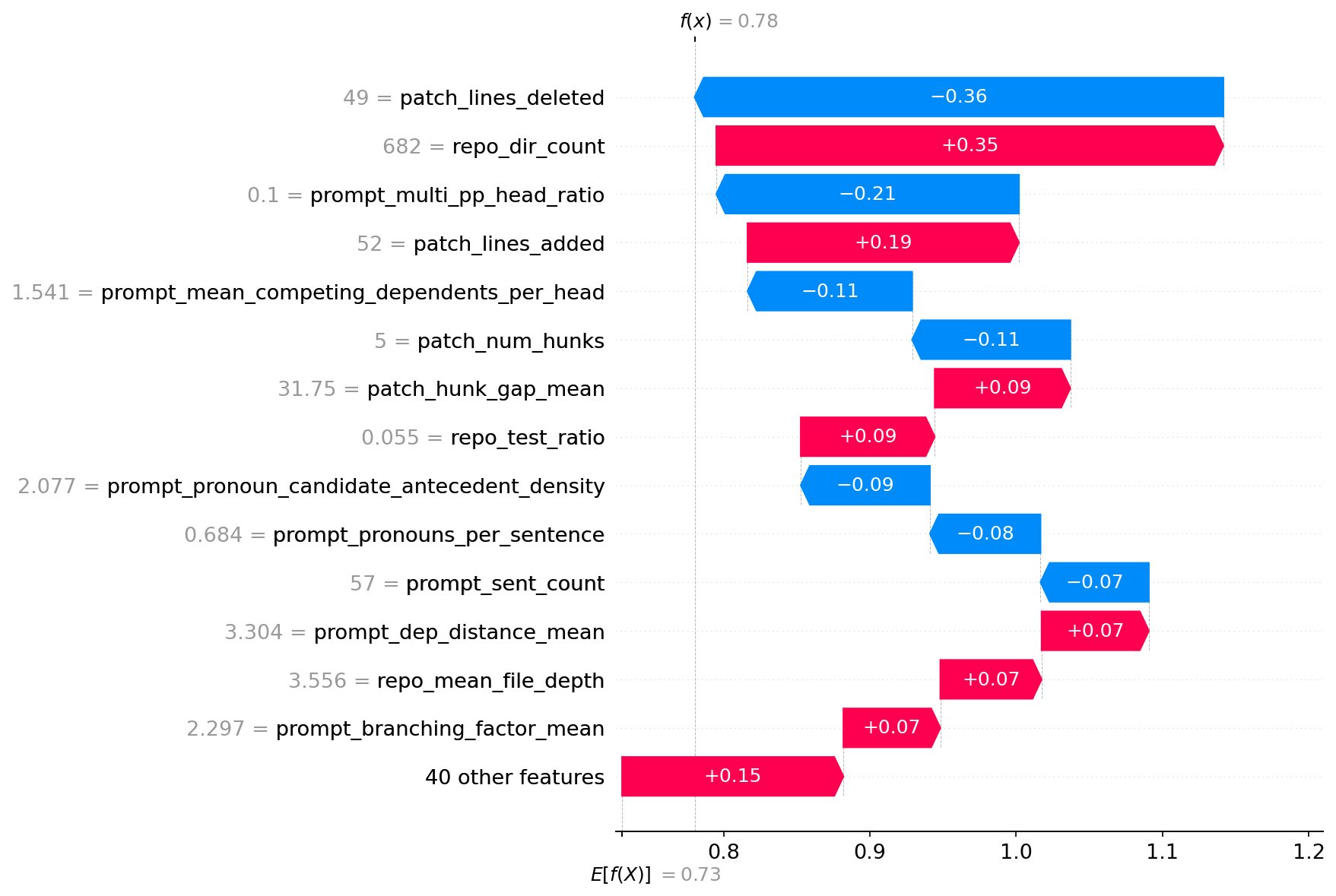}
    \caption{\texttt{faker}; $\hat{p} = 0.684$, observed: success.
      \texttt{prompt\_multi\_pp\_head\_ratio} ($-0.21$) is the third-largest
      contributor.}
    \label{fig:waterfall-avg-faker}
  \end{subfigure}
  \hfill
  \begin{subfigure}[t]{0.48\linewidth}
    \centering
    \includegraphics[width=\linewidth]{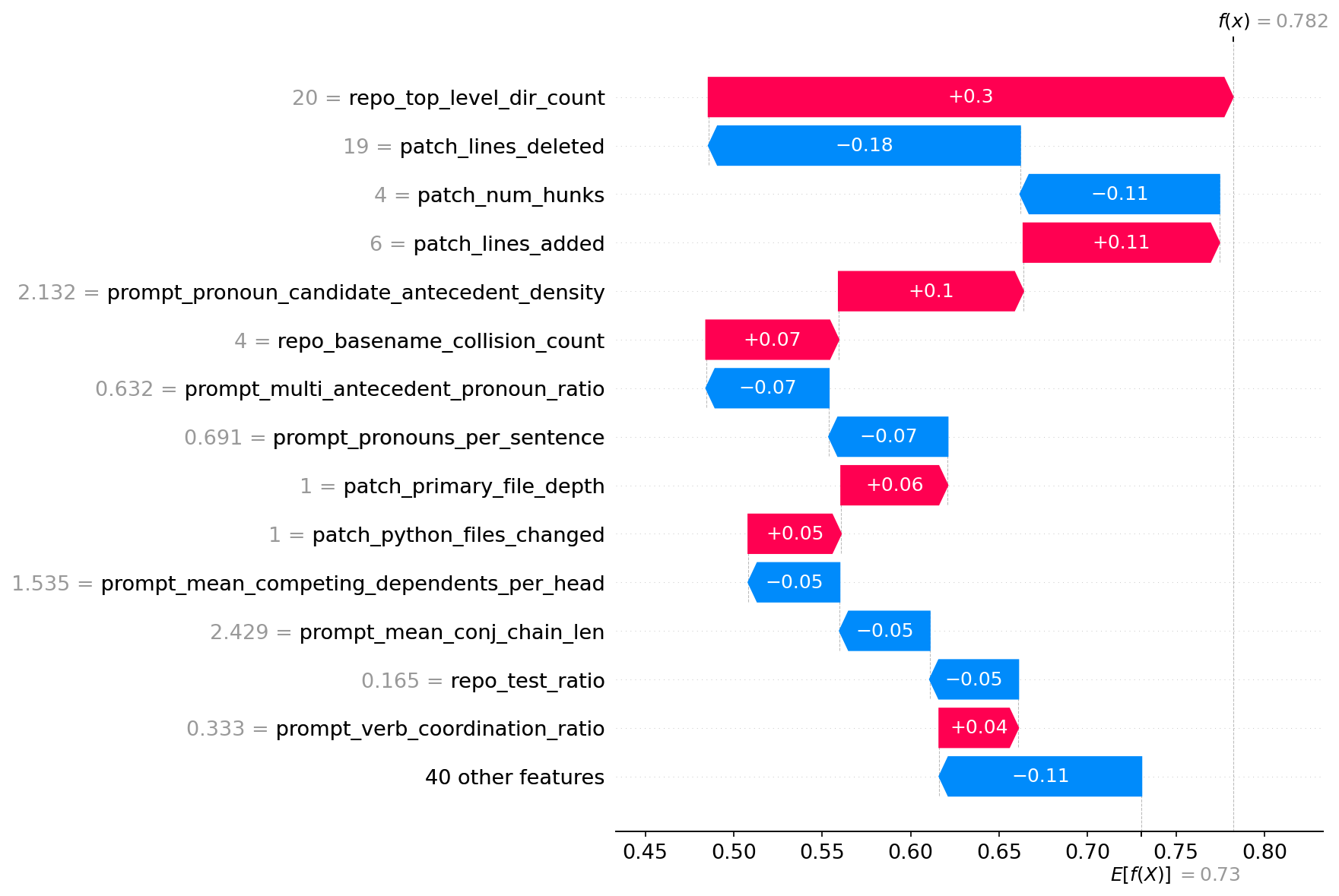}
    \caption{\texttt{parso}; $\hat{p} = 0.685$, observed: success.
      \texttt{prompt\_pronoun\_candidate\_antecedent\_density} ($+0.10$) ranks
      fifth among contributors.}
    \label{fig:waterfall-avg-parso}
  \end{subfigure}
  \caption{SHAP waterfalls for four representative average-difficulty tasks
    ($\hat{p} \approx 0.68$, $E[f(x)] = 0.73$, XGBoost, \texttt{any\_success}).
    In three of four cases a prompt feature ranks among the top contributors,
    with magnitudes comparable to the structural features for those tasks.}
  \label{fig:waterfalls-average}
\end{figure}

The waterfall plots (Figures~\ref{fig:waterfall-easiest}--\ref{fig:waterfall-hardest}) illustrate how these effects compose at the task level. 
All values are in log-odds space; the population baseline is $E[f(x)] = 0.73$. 
The easiest task ($\hat{p} = 0.997$, \texttt{pygments/pygments}) has a compact two-site edit: every structural contributor pushes the prediction upward from baseline. 
The hardest task ($\hat{p} = 0.013$, \texttt{getnikola/nikola}) has moderate patch volume but high \texttt{repo\_top\_level\_dir\_count} as the second-largest negative contributor---illustrating that the combination of moderate fragmentation and a wide repository suffices to drive the prediction to near-zero success. 
The average task ($\hat{p} \approx 0.68$, \texttt{kurtmckee/feedparser}, Figure~\ref{fig:waterfall-average}) sits just above baseline with mixed structural signals, a pattern examined further below.

\begin{figure}[!hbtp]
  \centering
  \includegraphics[width=\linewidth]{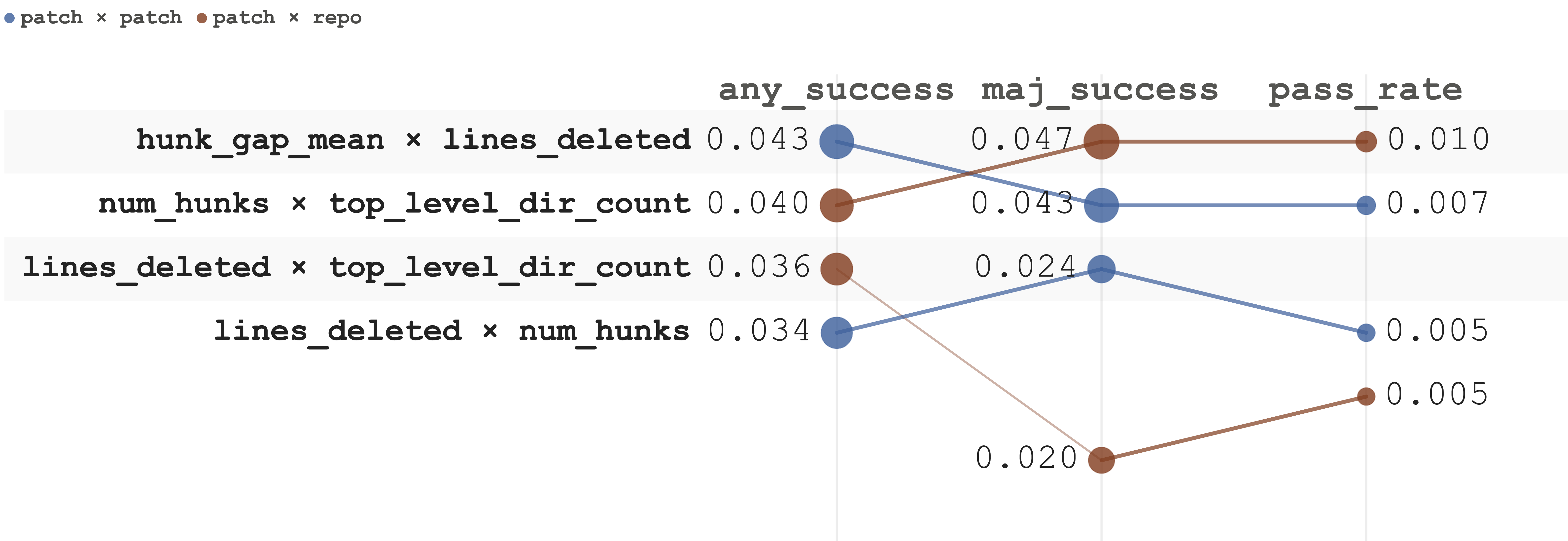}
  \caption{Rank--magnitude bump chart for the top four SHAP interaction pairs by mean absolute interaction value averaged across outcome targets. 
  }
  \label{fig:shap-interactions}
\end{figure}

Feature interactions are dominated by patch $\times$ repo pairs; Figure~\ref{fig:shap-interactions} shows the top four by mean absolute interaction value averaged across outcomes. 
The two leading pairs (\texttt{patch\_hunk\_gap\_mean} $\times$ \texttt{lines\_deleted} and \texttt{num\_hunks} $\times$ \texttt{top\_level\_dir\_count}) swap ranks between \texttt{any\_success} and the other two targets but remain the top interactions throughout. 

\subsubsection{Prompt Features in the Mid-Band}
At the population level, prompt features first appear at rank 16 and carry an order of magnitude less SHAP mass than the leading patch features (Figure~\ref{fig:shap-bump}), consistent with the RQ2 ablation. These aggregate results, however, can suppress changes in the relative importance of feature groups across the prediction range.

We stratify tasks by the predicted probability $\hat{p}$ for \texttt{any\_success}: \emph{easy} tasks fall in the top decile of $\hat{p}$ ($n{=}893$), \emph{hard} tasks in the bottom decile ($n{=}886$), and \emph{mid-band} tasks have predictions within $0.05$ of the mean prediction ($\overline{\hat{p}}{=}0.685$; $n{=}575$). This group has a median summed SHAP contribution is $+0.09$ log-odds, compared with $+2.84$ for easy tasks and $-3.76$ for hard tasks.

Prompt features become substantially more prominent in the mid-band. At least one prompt feature ranks among the five largest contributors for $\mathbf{70.3\%}$ of near-baseline tasks ($404/575$), $\mathbf{26.8\%}$ of easy tasks ($239/893$), and $\mathbf{6.8\%}$ of hard tasks ($60/886$). The same ordering holds for \texttt{pass\_rate} and \texttt{maj\_success}.

We observe that the gap between prompt and structural contributions begin to narrow here. We define
$[
R(x)=\frac{\max_{j\in\mathrm{prompt}}|\phi_j(x)|}
{\max_{j\in\mathrm{structural}}|\phi_j(x)|}
]$.
Median $R$ rises to $0.404$ near mid-band, compared with $0.205$  ($1.97\times$) for easy tasks and $0.108$ ($3.75\times$) for hard tasks. 

No single prompt feature accounts for the pattern. The three most frequent contributors are \texttt{prompt\_pronouns\_per\_sentence} (referential ambiguity, $11.1\%$ of near-baseline tasks; $64/575$), \texttt{prompt\_mean\_conj\_chain\_len} (coordination ambiguity, $10.1\%$; $58/575$), and \texttt{prompt\_mean\_competing\_dependents\_per\_head} (attachment ambiguity, $9.4\%$; $54/575$), all established markers of processing
difficulty in computational linguistics~\cite{gibson2000dependency}.

Figure~\ref{fig:waterfalls-average} illustrates this at the task level with three representative examples:

\texttt{prompt\_pp\_density\_per\_clause}, another attachment-ambiguity feature, is the second-largest contributor in a \texttt{sqlglot} task ($-0.20$); \texttt{prompt\_multi\_pp\_head\_ratio} ranks third in a \texttt{faker} task ($-0.21$); \texttt{prompt\_pronoun\_candidate\_antecedent\_density} ranks fifth in a \texttt{parso} task ($+0.10$).

\begin{tcolorbox}[
  colback=lipicsLightGray,
  colframe=lipicsLightGray,
  boxrule=0pt,
  arc=2pt,
  left=6pt, right=6pt, top=4pt, bottom=4pt
]

\textbf{Finding~3.}
The composition of task-difficulty signals changes across the prediction range. Patch fragmentation and repository scale dominate at the extremes, reliably separating easy from hard tasks across all three outcome definitions. Near the mid-band, prompt features enter the top-5 SHAP contributors for $70.3\%$ of tasks, versus $26.8\%$ (easy) and $6.8\%$ (hard). No single prompt feature displaces patch or repository features in aggregate importance.
\end{tcolorbox}

\section{Discussion}

Our results make the structural composition of a benchmark directly observable. 
This helps reason about whether two benchmarks are measuring the same thing or whether an improvement in solve rate reflects a genuine capability gain in targeted dimensions.

One practical application for our findings is in benchmark auditing~\cite{tu2026benchguard}. 
Computing difficulty scores for each task in an existing benchmark reveals whether the difficulty distribution is structurally balanced or concentrated in a narrow region of the task space. 
A benchmark heavily populated with compact, single-hunk patches in shallow repositories would systematically under-represent tasks that require broader navigation or multi-site coordination, which would be invisible in aggregate solve rates.
This can also provide opportunities for more intentional benchmark design, where tasks structure and difficulty are explicit considerations.

A second application area is in difficulty-stratified evaluation. An agent that improves primarily on tasks that are already easy for the reference agent represents a qualitatively different kind of progress from one that improves on structurally difficult tasks. 
Reporting performance within difficulty quartiles would make this distinction visible. 
The model produces well-calibrated probability estimates across the full range, so predicted probabilities can be used directly as continuous difficulty scores without post-hoc recalibration.

The features required for these applications are entirely deterministic. This is a meaningful practical advantage over IRT-based difficulty estimation approaches such as that of Ge et al.~\cite{ge2026agent}, which require LLM embeddings and rubric scoring per task.

\subsection{Toward Agent-Specific Difficulty Profiles}
The difficulty patterns we found are relative to a specific reference agent, and that framing opens a natural extension of the work. 
Applying our framework separately to different agents would make per-agent difficulty profiles observable. 
Prior work in software engineering prediction shows why this matters. 
Menzies et al.\ found that global rules learned across heterogeneous populations were consistently outperformed by rules learned from neighboring clusters~\cite{menzies2012local}, and Jiang et al.\ showed that developer-specific defect models substantially outperformed pooled ones because developers differ too much in experience and coding style to be well-described by a single model~\cite{jiang2013personalized}.
The same principle likely applies here. 
What is hard for Qwen3 with this scaffold (due to, for example, its approach to navigating wide directory structures) may be easier for an agent with a different localization strategy. 
As trajectory data accumulates across frontier agents, this form of per-agent difficulty diagnosis becomes a practical step toward understanding and improving the specific structural bottlenecks each system faces.

\subsection{Toward Calibrated Expectations in Development Tools}
Though not the focus of our study, our efforts provide valuable insights towards trustworthy agentic developer solutions. 
Our findings connect to how developers decide how much to trust an agent's output on a given task. 
According to the PICSE~\cite{10.1145/3772370} framework, trust in AI-assisted software tools is shaped in part by the expectations a developer brings to an interaction and by how well the system's reliability conditions are made explicit. 
A single aggregate benchmark score does not support this well as it hides the conditions under which the agent is more or less likely to succeed.

Difficulty scores give developers a more grounded basis for calibrating reliance. 
A developer who has a sense of how structurally complex a task is can look at an agent's performance within the corresponding difficulty band on a benchmark, rather than its overall solve rate, to form a more accurate expectation of whether the agent will succeed. 
This shifts the question from ``how good is this agent'' to ``how good is this agent on tasks like this one,'' which is the question that matters most for deciding when and how to use an AI-assisted tool.

\section{Threats to Validity}
\subsection{Internal}
\noindent\textbf{Agent pretraining on benchmark repositories.}
Qwen3-Coder-480B was pretrained on public code, and CoderForge-Preview draws from public GitHub repositories. Tasks from repositories present in pretraining may be solved at elevated rates unrelated to structural features.\\

\noindent\textbf{Outcome reliability and agent stochasticity.}
Pass rate is estimated from trajectories generated at nonzero temperature, so outcomes conflate genuine task difficulty with sampling variance intrinsic to the agent's decoding process. No static feature can predict this stochastic component, which partially explains the irreducible variance in the regression target.

\subsection{External}
\noindent\textbf{Single model group.}
All trajectories in CoderForge-Preview were generated by Qwen3-Coder-480B operating within an OpenHands scaffold. The features that predict difficulty for this agent may not generalize to agents with different architectures, context window sizes, or scaffolding strategies. In particular, our finding that repository scale and directory breadth are important predictors implicitly reflects this agent's navigation strategy; an agent with a different file-localization approach might be less sensitive to these features. Replication with trajectories from other frontier agents would strengthen generalizability claims.\\
 
\noindent\textbf{Python-centric sample.}
The underlying repositories are predominantly Python projects. Features related to Python file counts and test infrastructure reflect conventions of the Python ecosystem (e.g., \texttt{pytest}-based test organization). The absolute importance values for these features may not transfer directly to repositories in other languages, though the underlying constructs (test density, repository scale, edit fragmentation) are language-agnostic.

\subsection{Construct}
\noindent\textbf{Gold patch as ground truth.}
Patch features are computed from the gold (oracle) patch provided with each task. In practice, an agent generates a candidate patch that may differ structurally from the gold. A task could have a simple gold patch but require complex agent reasoning, or vice versa. We use the gold patch as the best available proxy for ``what a correct solution looks like,'' consistent with prior just-in-time defect prediction work. The strong predictive performance suggests this proxy is informative, but we acknowledge that gold-patch features do not capture all sources of task complexity.\\

\noindent\textbf{Prompt features.}
Our linguistic prompt features operationalize structural conditions for ambiguity rather than ambiguity itself. The RE literature distinguishes \emph{nocuous} ambiguity, where different readers reach genuinely different interpretations that lead to different implementations, from \emph{innocuous} ambiguity, where the structure is syntactically ambiguous but context resolves it to a single intended meaning~\cite{yang2010methodology}. Our features cannot make this distinction. A syntactically complex issue description may still be pragmatically unambiguous as the domain, the surrounding code context, or simple common sense may resolve the attachment uniquely. Conversely, a structurally simple sentence can be genuinely nocuous if the two readings lead to different code changes. The consequence is that our prompt features measure a necessary but not sufficient condition for the kind of specification ambiguity that would actually mislead an agent. This upper-bounds their predictive power and is consistent with the low global SHAP signal we observe for prompt features. Whether finer-grained ambiguity detection, for instance, using LLM-as-judge to classify issues as nocuous versus innocuous—would recover additional predictive signal is an open question for future work.

\section{Conclusion}
The difficulty of a software issue resolution task is encoded in its structure. 
We showed that static, deterministic features derived from the gold patch, repository, and issue prompt predict agent success with AUC $= 0.863$ without any model inference. 
This enables pre-hoc difficulty estimation and opens the door to structurally controlled benchmark construction, difficulty-stratified evaluation, and more grounded developer reliance on agent tools.
Future work will extend the framework across multiple agents to expose per-agent difficulty profiles and, in aggregate, a global difficulty landscape that characterizes the structural frontier of agentic software engineering. 
More broadly, the measurement framework generalizes naturally to other agentic SE tasks such as feature implementation, test generation, where analogous structural properties of the target artifact and the natural-language specification can be represented as static difficulty signals.

\section{Data Availability}
The replication package is available \href{https://github.com/ebtesam25/diffmodel-esem}{\textit{here}}.

\bibliography{camera-ready-esem}

\end{document}